\documentclass[a4paper,11pt]{article}
\pdfoutput=1

\usepackage{jheppub}
\usepackage[T1]{fontenc}
\usepackage{mathtools,amsmath,amsthm,amssymb,calrsfs}
\usepackage{dsfont,xcolor}

\DeclareMathAlphabet{\pazocal}{OMS}{zplm}{m}{n}

\title{\boldmath Supersymmetry of the static Reissner--Nordstr{\"o}m black hole in Bertotti--Robinson ($\mathrm{AdS}_2 \times \mathbb{S}^2$)}

\author[a,b]{Andrea Di Pinto}
\author[b]{and Adriano Vigan\`o}

\affiliation[a]{Dipartimento di Scienza e Alta Tecnologia (DISAT), Universit\`a degli Studi dell'Insubria \\
Via Valleggio 11, I-22100 Como, Italy}
\affiliation[b]{Istituto Nazionale di Fisica Nucleare (INFN), Sezione di Milano \\
Via Celoria 16, I-20133 Milano, Italy}

\emailAdd{andrea.dipinto@mi.infn.it}
\emailAdd{adriano.vigano@mi.infn.it}

\abstract{
We examine supersymmetry of charged and accelerating black holes embedded in a Bertotti--Robinson universe, in the context of $\pazocal{N}=2$, $D=4$ supergravity.
After a review of the solution, we study the constraints that guarantee supersymmetry and explicitly compute the Killing spinors of the spacetime.
We show that the supersymmetric solution saturates the BPS bound, and we use this result to compute the mass of the black hole and to analyze the thermodynamics of the solution.
Finally, we present a generalization of the extremal solution to include the cosmological constant.
}

\begin{document}

\maketitle

\flushbottom

\section{Introduction}
\label{sec-introduction}

Supergravity theories are recognized as the low-energy limit of string theories, and as such they represent a way to pave our understanding of M-theory and, more generally, quantum gravity.

Black holes are fundamental objects in our Universe, and play a prominent role in supergravity:
they are necessary, for instance, to compare the microscopic entropy computed as the ensemble of D-branes~\cite{Strominger:1996sh}, and the macroscopic entropy given by the Bekenstein--Hawking area law~\cite{Bekenstein:1973ur,Hawking:1975vcx,Hawking:1994ii}.
Moreover, since the discovery of the attractor mechanism~\cite{Ferrara:1995ih}, it has become crucial the construction and understanding of black hole solutions in supergravity theories.

Among exact solutions in supergravity, a prominent role is played by bosonic backgrounds preserving supersymmetry, since they are stable and less susceptible to quantum corrections.
Supersymmetric solutions are characterized by the existence of Killing spinors, which highly constrain the geometry of the spacetime:
for instance, Killing spinors define $G$-structures, which allow for the construction of bilinear tensors constructed out of such spinors~\cite{Gauntlett:2002sc,Gauntlett:2002nw}.
Furthermore, spinorial geometry is useful to classify supersymmetric solutions~\cite{Cacciatori:2007vn}.

The study of Killing spinors equations in $\pazocal{N}=2$, $D=4$ ungauged supergravity was initiated by Tod in~\cite{Tod:1983pm}, where he obtained a list of supersymmetric solutions using the Newman--Penrose formalism~\cite{Newman:1961qr}.
When the Killing vector obtained from the Killing spinor is timelike, the solution reduces to the Israel--Perjés--Wilson class~\cite{Perjes:1971gv,Israel:1972vx} which, as we will see below, is completely specified by a harmonic function defined on the three-dimensional flat base space.
The explicit study of supersymmetric charged black holes was pursued, e.g., by Romans~\cite{Romans:1991nq} in the static case, and then generalized to the stationary case in~\cite{Caldarelli:1998hg,Alonso-Alberca:2000zeh}.

Astrophysical black holes are typically surrounded by accretion disks that sustain strong magnetic fields~\cite{Eatough:2013nva,You:2023dax}.
Universes permeated by an external electromagnetic field have been known for a long time, with the Bonnor--Melvin~\cite{Bonnor:1954tis,Melvin:1963qx} and the Bertotti--Robinson~\cite{Bertotti:1959pf,Robinson:1959ev} solutions being the most notable examples:
despite being relativistic ``uniform'' electromagnetic spacetimes, they represent a good approximation of the behaviour of an (electro)magnetic field in an astrophysical setting.
Many exact solutions that describe black holes embedded in an electromagnetic field have been built over the years, starting with the pioneering works of Ernst and Wild~\cite{Ernst:1976mzr,Ernst:1976bsr}:
these kinds of solutions have been studied by many authors (see e.g~\cite{Thorne:1965xnn,Gibbons:2013yq,Booth:2015nwa}).

Recently, Ovcharenko and Podolsky found a family of spacetimes~\cite{Podolsky:2025tle,Ovcharenko:2025cpm,Ovcharenko:2025qov, Ovcharenko:2026byw,Ovcharenko:2026pow} describing black holes embedded in an electromagnetic Bertotti--Robinson universe.
Such solutions belong to the Petrov type D class and are part of the metrics\footnote{Such metrics are quite involved and their physical meaning has never been elucidated until the work of Ovcharenko and Podolsky.} found by Debever et al.~\cite{Debever:1983pi,Debever:1984yxe}.
Contrary to the Bonnor--Melvin spacetime, the Bertotti--Robinson universe does not create an infinitely strong electromagnetic field at infinity, thus it is more physically well behaved.

It is worth noting that Alekseev and Garcia also found black hole solutions of Petrov type D in a Bertotti--Robinson background~\cite{Alekseev:1996fq,Alekseev:2025czq}, but at the present moment it remains unclear whether they are, at least in part, equivalent to the static subcases of the Ovcharenko--Podolsky family~\cite{Ovcharenko:2026byw}.
Here, we focus on the Ovcharenko--Podolsky solutions.

The interest in black holes embedded in electromagnetic universes naturally raises the question of their supersymmetric extension within the framework of supergravity theories.
It is known that the Bonnor--Melvin spacetime itself is not supersymmetric, and as such the Ernst solutions do not admit Killing spinors.
On the other hand, it is known~\cite{Ferrara:1997yr} that the Bertotti--Robinson solution, being the near-horizon limit of an extremal Reissner--Nordstr{\"o}m black hole, which is a $\frac{1}{2}-$BPS solution, also admits Killing spinors and it is indeed maximally supersymmetric.
For this reason, there is hope that black holes in a Bertotti--Robinson background may admit Killing spinors.
Studying the supersymmetry of such solutions in $\pazocal{N}=2$, $D=4$ ungauged supergravity might be the first step to extend this class of solutions to the gauged version of the theory, along the lines of~\cite{Alonso-Alberca:2000zeh}.

The structure of the paper is as follows. In section~\ref{sec-review}, we review the new family of possibly accelerating and charged black hole solutions found in~\cite{Ovcharenko:2026byw}, highlighting the subcases that are of greatest interest to us. In section~\ref{sec-sugra}, we review the basic concepts of $\pazocal{N}=2$, $D=4$ supergravity, focusing on the supersymmetry conditions and the existence of Killing spinors in section~\ref{sec-killing}. We then specialize to the aforementioned class of black holes in section~\ref{sub-sec-killing-bh-br}, where we show that the only new supersymmetric subcase is the extremally charged Reissner--Nordstr{\"o}m black hole embedded in a Bertotti--Robinson universe, and derive the explicit expression for the Killing spinors of this spacetime. Building on the results of the last-mentioned section, we proceed to study the BPS bound of the extremal subcase in section~\ref{sec-BPS}, its thermodynamics in section~\ref{sec-thermo-extremal}, and a generalization with a cosmological constant in section~\ref{sec-cosmological-generalization}.

\section{Review of the solution}
\label{sec-review}

In this section we review the main properties of the solution describing a Reissner--Nordstr{\"o}m black hole accelerating in Bertotti--Robinson recently found by Ovcharenko and Podolsky in~\cite{Ovcharenko:2026byw}.

To set the conventions, we explicitly write down the Einstein--Maxwell equations
\begin{equation}
R_{\mu\nu} - \frac{1}{2} R\, g_{\mu\nu} =
2 \biggl( F_{\mu\rho} F_\nu^{\;\;\rho} - \frac{1}{4} F^2 g_{\mu\nu} \biggr) \,, \qquad
%\frac{1}{\sqrt{-g}} \partial_\mu \bigl(\sqrt{-g} F^{\mu\nu} \bigr) = 0 \,.
\nabla_\mu F^{\mu\nu} = 0 \,.
\end{equation}
The metric and the $1-$form that satisfy these field equations are given by
\begin{subequations}
\label{accelerating-rn-br}
\begin{align}
\label{accelerating-rn-br-metric}
{ds}^2 & = \frac{1}{\Omega^2}
\biggl[-\frac{\Delta_r} {r^2}{dt}^2 + r^2\biggl(\frac{{dr}^2}{\Delta_r}
+ \frac{{d\theta}^2}{\Delta_\theta}\biggr) + r^2\sin^2\theta \, \Delta_\theta \, {d\varphi}^2\biggr] \,, \\
\label{accelerating-rn-br-potential}
A & = \frac{w}{B} \frac{\Omega_{,\theta}}{r\sin\theta} dt
+ \frac{\sqrt{1-w^2}}{B} (\Omega - 1 - r\,\Omega_{,r}) d\varphi \,,
\end{align}
\end{subequations}
where
\begin{subequations}
\label{accelerating-rn-br-functions}
\begin{align}
\Delta_r & = \bigl[(r-r_0)^2\bigl(1-B^2 m^2\bigr)- 2 m (r-r_0)\bigr]\bigl[1-\bigl(\alpha^2-B^2\bigr)(r-r_0)^2\bigr] \,, \\
\Delta_\theta & = 1 - 2\alpha m \cos\theta + B^2 m^2 \cos^2\theta \,, \\
\begin{split}
\Omega^2 & = \bigl[1-\alpha(r-r_0)\cos\theta\bigl]^2 +\,\\
 &\quad + B^2 \bigl[(r-r_0)^2\bigl(\sin^2\theta+B^2m^2\cos^2\theta-2\alpha m \cos\theta\bigr)+ 2 m (r-r_0)\cos^2\theta\bigl] \,.
\end{split}
\end{align}
\end{subequations}
Two distinct \emph{discrete} subcases are considered\footnote{These two discrete branches are necessary to remove the twist from the more general class of black holes embedded in a Bertotti--Robinson background presented in~\cite{Ovcharenko:2025cpm}.} in~\cite{Ovcharenko:2026byw}:
\begin{subequations}
\begin{align}
\label{r0-zero}
\text{case I:} & \quad r_0=0 \,, \\
\label{r0-not-zero}
\text{case II:} & \quad r_0 = \frac{2 m B^2}{\alpha^2-B^2\bigl(1-B^2 m^2\bigr)} \,.
\end{align}
\end{subequations}
The first possibility, $r_0=0$ (case I), only contains \emph{uncharged} solutions, corresponding to a possibly accelerating Schwarzschild black hole with a Bertotti--Robinson electromagnetic field.
On the other hand, $r_0\neq0$ (case II), corresponds to an accelerating and charged Reissner--Nordstr{\"o}m black hole embedded in a Bertotti--Robinson electromagnetic universe.

We are interested in case II:
in this context, we can interpret the various parameters appearing in~\eqref{accelerating-rn-br}.
In particular, $B$ denotes the Bertotti--Robinson electromagnetic parameter, $w$ is the duality parameter that measures the dyonic charge ($w=1$ for a purely electric solution, $w=0$ for a purely magnetic solution), $\alpha$ is the acceleration parameter, and $m$ is related to the black hole mass.

Finally, the spacetime is affected by the presence of conical singularities along the axes $\theta=0$ and $\theta=\pi$, as can be checked by computing
\begin{align}
\delta_0 & = \lim_{\theta\to0} \frac{1}{\theta} \sqrt{\frac{g_{\varphi\varphi}}{g_{\theta\theta}}} = 2\pi (1 - 2m\alpha + B^2m^2) \,, \\
\delta_\pi & = \lim_{\theta\to\pi} \frac{1}{\pi-\theta} \sqrt{\frac{g_{\varphi\varphi}}{g_{\theta\theta}}} = 2\pi (1 + 2m\alpha + B^2m^2) \,.
\end{align}
The absence of nodal singularities is equivalent to the requirement $\delta_0=\delta_\pi=2\pi$:
if one of the two defects can be removed by choosing an appropriate periodicity for the coordinate $\varphi$, the other can be fixed only by removing the black hole ($m=0$), or by removing the acceleration parameter $\alpha=0$.

\subsection{Horizons and causal structure}
\label{subsec-horizons}

Before analyzing the supersymmetry properties of the spacetime~\eqref{accelerating-rn-br}, it is worth clarifying the nature of the event and acceleration horizons, depending on the ranges of the parameters.

The horizons are given by the roots of $\Delta_r$, which is a fourth-order polynomial in $r$.
Its discriminant is equal to $16 m^2(\alpha^2-B^2)\bigl[(1-B^2m^2)^2-4 m^2 (\alpha^2-B^2)\bigl]^2$, and thus its sign is controlled by the quantity $(\alpha^2-B^2)$.
Assuming that $\alpha$, $B$ and $m$ are all positive, we can distinguish among three cases.
\begin{itemize}
\item
$\alpha>B$:
the discriminant is positive or null, thus $\Delta_r$ has four real roots, given by
\begin{equation}
r_- = r_0 \,, \qquad
r_+ = r_0 + \frac{2m}{1-B^2m^2} \,, \qquad
r_{a\pm} = r_0 \pm \frac{1}{\sqrt{\alpha^2-B^2}} \,,
\end{equation}
which are all distinct when the discriminant is strictly positive, while $r_{+}$ can coincide with either $r_{a+}$ or $r_{a-}$ when the discriminant vanishes, i.e.~when $2m\alpha=1+B^2m^2$.
The ordering of the four roots is established by the ranges of the parameters:
more precisely, we notice that $1-B^2m^2<1$ always, therefore $r_0>0$ provided that $\alpha>B$.
Conversely, the relative positions of the horizons are determined by the specific values of the parameters, and thus different orderings are possible (however, it always holds that $r_{a-}<r_-<r_{a+}$):
%The possible orderings are as follows:
\begin{itemize}
\item
$r_{a-}<r_-<r_+<r_{a+}$:
when $Bm<1$ and $2m\alpha<1+B^2m^2$;
\item
$r_{a-}<r_-<r_+=r_{a+}$:
when $Bm<1$ and $2m\alpha=1+B^2m^2$;
\item
$r_{a-}<r_-<r_{a+}<r_+$:
when $Bm<1$ and $2m\alpha>1+B^2m^2$;
\item
$r_{a-}<r_+<r_-<r_{a+}$:
when $Bm>1$ and $2m\alpha<1+B^2m^2$;
\item
$r_{a-}=r_+<r_-<r_{a+}$:
when $Bm>1$ and $2m\alpha=1+B^2m^2$;
\item
$r_+<r_{a-}<r_-<r_{a+}$:
when $Bm>1$ and $2m\alpha>1+B^2m^2$.
\end{itemize}
Not all the horizons must be simultaneously present:
depending on the value of the parameters, $r_{a-}$ and/or $r_+$ might be negative, and in that case they are outside the chart (which is defined for $r>0$).
Moreover, depending on the relative positions, the horizons acquire different meanings:
$r_{a\pm}$ are interpreted as acceleration horizons, while $r_\pm$ are inner or outer horizons.

Since we are mainly interested in the extremal limit (described below in section~\ref{subsec-extremal-rn-br}), the more relevant causal structures are the first and the fourth one:
they have a continuous limit in which the inner and the outer horizons collapse to a single horizon.
\item 
$\alpha=B$:
the discriminant is null, thus there are two pairs of equal real roots given by
\begin{equation}
r_- = r_0 \,, \qquad
r_+ = \frac{r_0}{1-B^2m^2} \,.
\end{equation}
Here $r_0=\frac{2}{B^2m}>0$.
Since $1-B^2m^2<1$, it holds $r_+>r_-$ for any value of the parameters:
they represent an outer (event) horizon and an inner (Cauchy) horizon.
\item 
$\alpha<B$:
the discriminant is negative, thus there are two real roots (and two complex roots)
\begin{equation}
r_- = r_0 \,, \qquad
r_+ = r_0 + \frac{2m}{1-B^2m^2} \,.
\end{equation}
In this case no acceleration horizon is present, and $r_0$ can be positive or negative, according to the value of the parameters.
\end{itemize}

\subsection{Bertotti--Robinson universe in accelerating coordinates}
\label{subsec-acc-rn}

For our purposes in what follows, it is useful to consider the subcase $m=0$ within the $r_0=0$ branch (case I) of the latter solution, from which one obtains
\begin{equation}
\label{acc-br}
\Delta_r = r^2\bigl[1-\bigl(\alpha^2-B^2\bigr)r^2\bigr] \,, \qquad
\Delta_\theta = 1 \,,\qquad
\Omega^2 = \bigr(1 - \alpha r \cos\theta\bigl)^2 + B^2 r^2\sin^2\theta \,.
\end{equation}
This metric is the conformally flat Bertotti--Robinson spacetime expressed in uniformly accelerating coordinates, in the sense that the acceleration parameter $\alpha$ can be removed from this solution by means of a coordinate transformation, as shown in~\cite{Ovcharenko:2026byw}.

\subsection{Extremal Reissner--Nordstr{\"o}m black hole in Bertotti--Robinson}
\label{subsec-extremal-rn-br}

Another subcase of particular interest for us is the extremal subcase of the previous general solution.
It is obtained by redefining the mass parameter $m \mapsto \frac{m(\alpha^2-B^2)}{2B^2}$, and then taking the limit $\alpha \rightarrow B$ in the $r_0\neq0$ branch (case II), which gives:
\begin{equation}
\label{extremal-rn-br}
\Delta_r = (r-m)^2 \,, \qquad
\Delta_\theta = 1 \,,\qquad
\Omega^2 = 1 - 2 B(r-m)\cos\theta + B^2(r-m)^2 \,.
\end{equation}
In this case no acceleration horizon is present and there are no conical singularities.

The resulting subcases of this solution are straightforward:
for $B=0$ we obtain the usual extremal Reissner--Nordstr{\"o}m black hole, while for the $m=0$ it reduces to the Bertotti--Robinson background in uniformly accelerating coordinates~\eqref{acc-br} where the acceleration parameter is equal to the Bertotti--Robinson one, $\alpha=B$.

\section{\texorpdfstring{$\pazocal{N}=2$}{N=2}, \texorpdfstring{$D=4$}{D=4} gauged supergravity}
\label{sec-sugra}

$\pazocal{N}=2$ supergravity in $D=4$ spacetime dimensions is a theory characterized by four bosonic and four fermionic degrees of freedom~\cite{Andrianopoli:1996cm}:
a graviton $e_\mu^a$, two gravitini $\bigl(\psi_\mu^1,\psi_\mu^2\bigr)$ which can be combined into a single complex spinor $\psi_\mu=\psi_\mu^1+\psi_\mu^2$ and a Maxwell gauge field $A_\mu$ minimally coupled to the gravitini with strength $1/\ell$.
The gauged version of the theory also includes a \emph{negative} cosmological constant, fixed by supersymmetry to be $\Lambda=-3/\ell^2<0$, and is described by the following Lagrangian:
\begin{equation}
\label{susy-lagrangian}
\begin{split}
e^{-1}\mathcal{L} & = -\frac{1}{4}R  - \frac{3}{2\ell^2}  + \frac{1}{4}F_{\mu \nu}F^{\mu \nu} + \frac{1}{2}\bar{\psi}_\mu\Bigl(\gamma^{\mu \nu \rho}\hat{\pazocal{D}}_\nu - \frac{1}{\ell}\gamma^{\mu \rho}\Bigr)\psi_\rho \\
\quad &  + \frac{i}{8}\bigl(F^{\mu \nu} + \hat{F}^{\mu \nu}\bigr) \bar{\psi}_\rho\gamma_{[\mu }\gamma^{\rho\sigma}\gamma_{\nu ]}  \psi_\sigma \,, 
\end{split}
\end{equation}
where $\hat{\pazocal{D}}_\mu$ denotes the gauge and Lorentz--covariant derivative
\begin{equation}
\label{gaugecovder}
\hat{\pazocal{D}}_\mu = \pazocal{D}_\mu - \frac{i}{\ell}A_\mu \,,
\end{equation}
$\pazocal{D}_\mu$ is the Lorentz--covariant derivative
\begin{equation}
\pazocal{D}_\mu = \partial_\mu + \frac{1}{4}{\omega_\mu}^{ab}\gamma_{ab} \,,
\end{equation}
while $\hat{F}_{\mu \nu}$ corresponds to the supercovariant field strenght
\begin{equation}
\label{supcovfs}
\hat{F}_{\mu \nu} = F_{\mu \nu} - \text{Im}(\bar{\psi}_\mu\psi_\nu) \,.
\end{equation}
The theory is given in a first-order formalism for the gravity sector, which means that the scalar curvature $R$ is a function of both the tetrad $e_{\mu}^{a}$ and the spin connection $\omega_{\mu}^{\;\;ab}$.
Hence, the spin connection is fixed by its own algebraic equations of motion following from the Lagrangian~\eqref{susy-lagrangian}, which are
\begin{equation}
\omega_{\mu ab} = \Omega_{\mu ab} - \Omega_{\mu ba} - \Omega_{ab\mu} \,,
\end{equation}
where
\begin{equation}
\Omega_{\mu\nu}^{\;\;\;\;a} \coloneqq \partial^{}_{[\mu} e^{a}_{\nu]} - \frac{1}{2} \text{Re} (\bar{\psi}_\mu\gamma^a\psi_\nu) \,.
\end{equation}
The action corresponding to the Lagrangian~\eqref{susy-lagrangian} is invariant under the following $\pazocal{N}=2$ local supertransformations
\begin{equation}
\label{transfsusy}
\delta e_\mu^{a} = \text{Re}(\bar{\epsilon}\gamma^a\psi_\mu) \,, \qquad
\delta A_\mu = \text{Im}(\bar{\epsilon}\psi_\mu) \,, \qquad
\delta \psi_\mu = \hat{\nabla}_\mu \epsilon \,,
\end{equation}
where $\epsilon$ is an infinitesimal Dirac spinor, and $\hat{\nabla}_\mu$ is the supercovariant derivative given by
\begin{equation}
\hat{\nabla}_\mu = \hat{\pazocal{D}}_\mu + \frac{1}{2\ell}\gamma_\mu + \frac{i}{4}\hat{F}_{ab}\gamma^{ab} \gamma_\mu \,. \label{supcovder}
\end{equation}
Furthermore, the supersymmetry algebra of gauged $\pazocal{N}=2$, $D=4$ supergravity is osp$(4|2)$.
This algebra has the ten bosonic generators $M_{ab}$, $M_{a4}$ ($a=0,1,2,3$) of the AdS subalgebra so$(3,2)$, two fermionic generators $Q_{\alpha}^n$ ($n=1,2$), plus one additional bosonic generator of SO${(2)}$ transformations, rotating the two supersymmetries into each other.
The basic anticommutator is then given by
\begin{equation}
\Bigl\{Q^n_{\alpha},Q^m_{\beta}\Bigr\} = \delta^{nm}\Bigl(\bigl(\gamma^a M_{a4} + i\gamma^{ab}M_{ab}\bigr) C\Bigr)_{\alpha\beta} + i \Bigl (C_{\alpha\beta}\mathrm{Q} + i \bigl(C\gamma^5\bigr)_{\alpha\beta}\mathrm{P} \Bigl)\varepsilon^{nm} \,,
\end{equation}
wherein $\mathrm{Q}$ and $\mathrm{P}$ are the electric and magnetic central charges, respectively;
$C$ denotes the charge conjugation matrix;
and $\varepsilon^{nm}$ is the permutation symbol in two dimensions.

In what follows, we consider the ungauged theory and thus set $\ell\to\infty$, corresponding to the absence of a cosmological constant, $\Lambda = 0$.

\section{Supersymmetry conditions and Killing spinors}
\label{sec-killing}

In the bosonic sector of this theory, $\psi_\mu=0$, the field equations following from~\eqref{susy-lagrangian} coincide with the Einstein--Maxwell equations with a negative cosmological constant.
Hence, spacetimes that satisfy these field equations represent possible background solutions of gauged $\pazocal{N}=2$ supergravity.
Furthermore, being $\psi_\mu=0$, the invariance of these background solutions under the supertransformations~\eqref{transfsusy} results in the equation
\begin{equation}
\label{killing-spinor}
\hat{\nabla}_\mu \epsilon = 0 \,.
\end{equation}
A spinor $\epsilon$ that satisfies the condition given by equation~\eqref{killing-spinor} is called a Killing spinor.

The basic integrability conditions for~\eqref{killing-spinor} reads
\begin{equation}
\label{integr-cond}
\hat{R}_{\mu \nu}\,\epsilon = 0 \,,
\end{equation}
where
\begin{equation}
\label{supercurv}
\hat{R}_{\mu \nu} = \bigl[\hat{\nabla}_\mu , \hat{\nabla}_\nu\bigr] \,,
\end{equation}
is referred to as the supercurvature.
Moreover, the integrability conditions for Killing spinors~\eqref{integr-cond} are also equivalent to
\begin{equation}
\label{integr-cond-2}
\det\bigl(\hat{R}_{\mu \nu}\bigr) = 0 \,.
\end{equation}
It is important to note that the basic integrability conditions~\eqref{integr-cond} are necessary, but not sufficient for the existence of Killing spinors.
They ensure that a Killing spinor exists locally, but there may be topological reasons that prevent their global existence.
For this reason, one must either consider a sufficient set of higher integrability conditions~\cite{vanNieuwenhuizen:1983wu} or return to the original differential equation~\eqref{killing-spinor}.

\subsection{Integration of the Killing spinors}
\label{sub-sec-killing-bh-br}

Given the general static solution representing an accelerating Reissner--Nordstr{\"o}m black hole in a Bertotti--Robinson universe~\eqref{accelerating-rn-br}, we can compute the integrability conditions~\eqref{integr-cond} for the two branches distinguished by different values of the $r_0$ parameter.
In the $r_0=0$ branch (case I), we have that the conditions can be satisfied only if $m=0$, which corresponds to the Bertotti--Robinson background in a uniformly accelerating coordinate system~\eqref{acc-br};
similarly, in the $r_0\neq0$ branch (case II), the conditions require the Reissner--Nordstr{\"o}m black hole to be extremal~\eqref{extremal-rn-br}.

However, as discussed in section~\ref{sec-introduction}, for the Bertotti--Robinson background in the abscence of the black hole, the acceleration parameter $\alpha$ is just an artifact of the coordinate system, which means that the extremal subcase~\eqref{extremal-rn-br} for $r_0\neq0$ already represents the most general static solution of this class that satisfies the integrability conditions~\eqref{integr-cond}.
Therefore, without loss of generality, we look for the Killing spinors of the extremal Reissner--Nordstr{\"o}m-Bertotti--Robinson black hole.

To solve the Killing equation~\eqref{killing-spinor}, we notice that the determinant of the connection part of the $t-$derivative, i.e.~$M_t\coloneqq\frac{1}{4}{\omega_t}^{ab}\gamma_{ab}+\frac{i}{4}\hat{F}_{ab}\gamma^{ab} \gamma_t$, is zero.
This means that there exists a non-trivial solution to the equation $\hat{\nabla}_t\epsilon=\partial_t\epsilon=0$, and such a solution is the combination of the eigenvectors of the matrix $M_t$.
Moreover, the eigenvectors will not depend on $t$ because of the latter equation.
The Killing spinors we are looking for thus have the following structure:
\begin{equation}
\epsilon(r,\theta,\varphi) =
c_+(r,\theta,\varphi)
\begin{pmatrix}
i \sqrt{1-w^2} \\
i w \\
1 \\
0
\end{pmatrix}
+
 c_-(r,\theta,\varphi)
\begin{pmatrix}
- i w  \\
i \sqrt{1-w^2} \\
0 \\
1
\end{pmatrix}
\,.
\end{equation}
Direct integration of the supersymmetry equations~\eqref{killing-spinor} yields
\begin{subequations}
\begin{align}
c_+ & = -Y_{+}\bigl(c_{1} X_{-}+c_{2} X_{+}\bigr) + Y_{+}\bigl(c_{1} X_{+}+c_{2} X_{-}\bigr)w + Y_{-}\bigl(c_{1} X_{-}-c_{2} X_{+}\bigr)\sqrt{1-w^2} \,, \\
c_- & =  -Y_{-}\bigl(c_{1} X_{+}+c_{2} X_{-}\bigr) - Y_{-}\bigl(c_{1} X_{-}-c_{2} X_{+}\bigr)w + Y_{+}\bigl(c_{1} X_{+}+c_{2} X_{-}\bigr)\sqrt{1-w^2} \,,
\end{align}
\end{subequations}
where we defined
\begin{subequations}
\begin{align}
X_\pm & \coloneqq \kappa(r,\theta)\Bigl(e^{-\frac{i \theta}{2}}\pm e^{\frac{i \theta}{2}}\Bigr) \bigl[1\mp B(r-m)\bigr] \,,\\
Y_\pm & \coloneqq e^{-\frac{i \varphi}{2}}\pm e^{\frac{i \varphi}{2}}\,, \\
\kappa & \coloneqq \frac{\Delta_{r}^\frac{1}{4}}{32^{\frac{1}{2}} r^{\frac{1}{2}} \Omega^{\frac{3}{2}}}= \frac{\sqrt{1-\frac{m}{r}}}{4\sqrt{2}\bigl[1-2 B \cos\theta(r-m)+B^2(r-m)^2\bigr]^{\frac{3}{4}}} \,.
\end{align}
\end{subequations}
Therefore, we have two independent Killing spinors, whose explicit expression is given by
\begin{subequations}
\label{killing-spinor-extremal}
\begin{align}
\epsilon_1(r,\theta,\varphi) & = c_1
\begin{pmatrix}
-i\, Y_{+}X_{-}\sqrt{1-w^2}+i\,Y_{-}(X_{+} w + X_{-}) \\
-i\, Y_{-}X_{+}\sqrt{1-w^2}-i\,Y_{+}(X_{-} w - X_{+}) \\
Y_{-}X_{-}\sqrt{1-w^2}+Y_{+}(X_{+} w - X_{-}) \\
Y_{+}X_{+}\sqrt{1-w^2}-Y_{-}(X_{-} w + X_{+}) \,
\end{pmatrix}
\,, \\
\epsilon_2(r,\theta,\varphi) & = c_2
\begin{pmatrix}
-i\, Y_{+}X_{+}\sqrt{1-w^2}-i\,Y_{-}(X_{-} w + X_{+}) \\
i\, Y_{-}X_{-}\sqrt{1-w^2}-i\,Y_{+}(X_{+} w - X_{-}) \\
-Y_{-}X_{+}\sqrt{1-w^2}+Y_{+}(X_{-} w - X_{+}) \\
Y_{+}X_{-}\sqrt{1-w^2}+Y_{-}(X_{+} w + X_{-})
\end{pmatrix}
\,.
\end{align}
\end{subequations}
Since we have two independent complex Killing spinors, the solution is $\frac{1}{2}$-BPS.
The current associated to such spinors
\begin{equation}
\mathrm{J}^{\mu}\partial_{\mu}=\sum_{n,m = 1}^2\bar{\epsilon}_{n}\gamma^{\mu}\epsilon_{m}\partial_{\mu}=-\Bigl[\bigl(|c_{1}|^2+|c_{2}|^2\bigr)-\bigl(c_{1}^{*} c_2 + c_{2}^{*} c_{1}\bigr)w\Bigr]\partial_{t} \,,
\end{equation}
is timelike, since
\begin{equation}
g_{\mu \nu}\,\mathrm{J}^{\mu}\mathrm{J}^{\nu} = -\frac{\Bigl[\bigl(|c_{1}|^2+|c_{2}|^2\bigr)-\bigl(c_{1}^{*} c_2 + c_{2}^{*} c_{1}\bigr)w\Bigr]\bigl(1-\frac{m}{r}\bigr)^2}{\bigr[1-2 B \cos\theta(r-m)+B^2(r-m)^2\bigl]} \,.
\end{equation}
%\begin{equation}
%\mathrm{J}^{\mu}\mathrm{J}_{\mu} = -\frac{\Bigl[|c_{1}|^2+|c_{2}|^2-\bigl(c_{1}^{*} c_2 + c_{2}^{*} c_{1}\bigr)w\Bigr]^2\Delta}{r^2\Omega^2}
%\end{equation}

\section{BPS bound and Majumdar--Papapetrou spacetime}
\label{sec-BPS}

Provided that the spacetime preserves half of the supersymmetries, it is natural to wonder if the black hole saturates the BPS bound~\cite{Witten:1978mh}, i.e.~if
\begin{equation}
\label{bps-bound}
\mathrm{M}^2 \geq |\pazocal{Z}|^2 \,,
\end{equation}
is satisfied with the equality.

The central charge is given by\footnote{One can find the central charge for the Einstein--Maxwell theory from the general action for $\pazocal{N}=2$, $D=4$ matter-coupled supergravity~\cite{Andrianopoli:1996cm} by choosing the prepotential $X^0=1/2$.}
$\pazocal{Z}=\mathrm{Q}+i\mathrm{P}$, and in particular
\begin{equation}
\label{charges-extremal}
\mathrm{Q} = -\frac{1}{8\pi}\int_\Sigma F^{\mu\nu} dS_{\mu \nu} = m w \,, \qquad
\mathrm{P} = -\frac{1}{8\pi}\int_\Sigma {}^\star F^{\mu\nu} dS_{\mu \nu} = m \sqrt{1-w^2} \,,
\end{equation}
where ${}^\star F^{\mu\nu} = \frac{1}{2\sqrt{-g}}\varepsilon^{\mu \nu \rho \sigma} F_{\rho \sigma}$ is the Hodge dual of the Faraday tensor $F^{\mu \nu}$.
Equation~\eqref{charges-extremal} shows that for $m=0$, i.e.~in the absence of the black hole, the charges are zero as expected, since the background has no monopoles (consistent with the results of~\cite{Hu:2026slp}).

If the central charge can be easily found, it is not clear how to compute the total mass of the spacetime because of the asymptotics:
the spacetime is not asymptotically flat, so the ADM and Komar formalisms are not applicable;
the Komar integral, in particular, leads to a divergent result.
One might appeal to a surface charge computation, however even in that case it is not clear which form the mass generator at infinity should take, thus we lack of a direct calculation~\cite{Hu:2026slp}.
We can, however, make use of the Majumdar--Papapetrou form of the metric~\eqref{extremal-rn-br} to deduce the mass in the extremal case.

Indeed, we can introduce cylindrical coordinates for the extremal Reissner--Nordstr{\"o}m-Bertotti--Robinson black hole~\eqref{extremal-rn-br}
\begin{subequations}
\begin{align}
\rho & = \frac{\sin\theta\sqrt{\Delta_r \Delta_\theta}}{\Omega^2} = \frac{(r-m)\sin\theta}{\bigl[1 + B^2(r-m)^2 -2 B(r-m)\cos\theta\bigl]} \,, \\
z & = \frac{\sqrt{\Delta_r}\, \cos\theta - B^2 \Delta_r}{\Omega^2} = \frac{(r-m) \cos\theta - B^2 (r-m)^2}{\bigl[1 + B^2(r-m)^2 -2 B(r-m)\cos\theta\bigl]} \,,
\end{align}
\end{subequations}
to find
\begin{subequations}
\label{double-bertotti-cyl}
\begin{align}
ds^2 & = -U^{-2} {dt}^2 + U^2 \bigl( {d\rho}^2 + {dz}^2 + \rho^2 {d\varphi}^2 \bigr) \,, \\
A & = w \, U^{-1} dt + \sqrt{1-w^2} \, V d\varphi \,,
\end{align}
\end{subequations}
with functions
\begin{subequations}
\begin{align}
U & = \frac{m}{\sqrt{\rho^2+z^2}} + \frac{1}{B\sqrt{\rho^2+\bigl(z+\frac{1}{B}\bigr)^2}} \,, \\
V & = \frac{m z}{\sqrt{\rho^2+z^2}} + \frac{z+\frac{1}{B}}{B\sqrt{\rho^2+\bigl(z+\frac{1}{B}\bigr)^2}} -\frac{1}{B}\,.
\end{align}
\end{subequations}
The extremal solution thus belongs to the Majumdar--Papapetrou class~\cite{Majumdar:1947eu,Papaetrou:1947ib}.
To interpret this result, we can further introduce Cartesian coordinates $x=\rho\cos\varphi$, $y=\rho\sin\varphi$, and rewrite the solution in the more familiar form
\begin{subequations}
\label{mp-cartesian}
\begin{align}
ds^2 & = -U^{-2}{dt}^2 + U^2 \bigl( {dx}^2 + {dy}^2 + {dz}^2 \bigr) \,,\\
A & = w \, U^{-1} dt + \sqrt{1-w^2}  \frac{V}{x^2+y^2} (xdy - ydx) \,,
\end{align}
\end{subequations}
where now
\begin{equation}
U = \frac{m}{|\vec{x}|} + \frac{1}{B|\vec{x}-\vec{x}_B|} \,, \qquad
V = \frac{m z}{|\vec{x}|} + \frac{z-z_B}{B|\vec{x}-\vec{x}_B|} - \frac{1}{B} \,,
\end{equation}
and we defined the vectors $\vec{x}=(x,y,z)$ and $\vec{x}_B=(0,0,-1/B)$.
Our solution differs from the most known form of the Majumdar--Papapetrou solution, which indeed represents a collection of $N$ charged extremal black holes, in the absence of the unit term in the definition of $U$.

From the last expression, we can interpret the extremally charged black hole embedded in the Bertotti--Robinson universe as a collection of two sources, one located at the origin of the axes and with mass $m$, and the other located at the point $\vec{x}_B$ and with mass equal to the position along the $z-$axis $1/B$.
This picture is consistent with the Bertotti--Robinson spacetime as the near-horizon limit of the extremal Reissner--Nordstr{\"o}m black hole.

The natural generalization of the solution~\eqref{mp-cartesian} to include $N$ sources is obtained by choosing
\begin{equation}
U = \sum_{i=1}^N \frac{h_i}{|\vec{x}-\vec{x}_i|} \,,
\end{equation}
where $h_i$ denote the sources and $\vec{x}_i$ the positions of the sources.
Such a solution is known in the literature:
it represents the near-horizon limit of $N$ Reissner--Nordstr{\"o}m black holes~\cite{Maldacena:1998uz}.

The general form contains a collection of $N$ generic sources, which can be chosen at will to represent a black hole of mass $m_i=h_i$ or a Bertotti--Robinson field of strength $1/B_i=h_i$ (i.e.~the near-horizon limit of a charged hole) in various positions.
At first sight, the freedom to interpret a source as a black hole or as a magnetic field may be disorienting, however it fits into the equivalence principle:
indeed, once we have chosen the role of the sources, it is possible to perform a change of coordinates (a translation plus a rescaling), to convert a hole into a field, and vice versa.
Physically, a change of coordinates corresponds to a change in the observer, thus it makes perfect sense that by translating the position of the observer and rescaling her ruler, it is possible to zoom in or zoom out the hole, thus observing a ``small'' hole (the source $m_i$) or a ``large'' hole (the field $1/B_i$).

The form~\eqref{mp-cartesian} is particularly suitable to discuss the BPS bound~\eqref{bps-bound}.
According to the theorem proved in~\cite{Gibbons:1982fy}, the only solution to the Einstein--Maxwell equations saturating the BPS bound~\eqref{bps-bound} whose supercovariantly constant spinor $\epsilon$ is time-independent and whose current vector $\mathrm{J}^{\mu}=\bar{\epsilon}\gamma^\mu\epsilon$ is timelike, must belong to the Majumdar--Papapetrou metrics.\footnote{One can express our metric in the form given in~\cite{Gibbons:1982fy} by setting $U=W$ and $w=\cos\theta$.}
Then, the total mass of the spacetime is given by
\begin{equation}
\label{mass}
\mathrm{M} = |\pazocal{Z}| = m \,,
\end{equation}
which is the expected result.

It is worth noticing that the computation of the total mass cannot be pursued by means of a Komar integral, which gives a divergent result, nor by a phase space computation since, as stressed above and in~\cite{Hu:2026slp}, it is not clear how to choose the proper normalization for the timelike generator.
This makes the result~\eqref{mass} precious, being the only accessible way to compute the total mass of the (extremal) black hole, and shows the power of supersymmetry techniques in the investigation of black hole physics.

As a further comment on the Majumdar--Papapetrou metric~\eqref{double-bertotti-cyl}, we have that
\begin{equation}
\lim_{B\rightarrow0}\frac{1}{B\sqrt{\rho^2+\bigl(z+\frac{1}{B}\bigr)^2}} = 1 \qquad
\Rightarrow \qquad
\lim_{B\rightarrow0} U = 1 + \frac{m}{\sqrt{\rho^2+z^2}} \,,
\end{equation}
which is the single Reissner--Nordstr{\"o}m black hole written in cylindrical form.
On the other hand, for $m=0$ we obtain the Bertotti--Robinson spacetime in cylindrical coordinates.

\section{Thermodynamics of the extremal solution}
\label{sec-thermo-extremal}

We can evaluate the thermodynamic quantities in the extremal case\footnote{In a similar way, the general case is discussed in appendix~\ref{app-thermo-charges-general}.} in order to check the validity of the Smarr formula~\cite{Smarr:1972kt} and the first law of black hole mechanics~\cite{Bardeen:1973gs}, using the mass derived from the saturation of the BPS bound~\eqref{mass}.

In particular, the horizon area $\mathrm{A}$ and the Bekenstein--Hawking entropy $\mathrm{S}$~\cite{Bekenstein:1973ur,Hawking:1975vcx} are
\begin{equation}
\mathrm{A} = \int_{0}^{2 \pi}d\varphi \int_{0}^{\pi} \sqrt{g_{\varphi \varphi}g_{\theta\theta}}\Bigg\rvert_{r=m} \! = 4 \pi m^2 \,, \qquad
\mathrm{S} = \frac{\mathrm{A}}{4} = \pi \mathrm{M}^2\,, \label{entropy-extremal}
\end{equation}
while the surface gravity of the horizon $\kappa$ and the corresponding temperature $\mathrm{T}$ vanish, as expected since the black hole is extremal
\begin{equation}
\kappa = \sqrt{-\frac{1}{2} \nabla_{\mu}\xi_{\nu}\nabla^{\mu}\xi^{\nu}}\Bigg\rvert_{r=m} \! = 0 \,, \qquad
\mathrm{T} =\frac{\kappa}{2\pi} = 0 \,, \label{temperature-extremal}
\end{equation}
where $\xi = \partial_t$ is the timelike Killing vector. Finally, the electrostatic potential $\Phi$ is given by
\begin{equation}
\Phi_{H} = -A_{\mu}\xi^{\mu}\Big\rvert_{r=m} \! = 0\,, \qquad \qquad \Phi = \Phi_{H}+\Phi_{int}=\Phi_{int}, \label{electromagnetic-potential-extremal}
\end{equation}
where $\Phi_{int}$ is a gauge constant, which can always be set to zero for an asymptotically flat spacetime. However, since the solution under consideration is not asymptotically flat, we leave it free and, if necessary, subsequently fix it in order to satisfy the thermodynamic laws.
Indeed, by setting $w=1$ to eliminate the magnetic component of the electromagnetic field, it follows that both the Smarr law
\begin{equation}
\mathrm{M} = 2 \,\mathrm{T}\,\mathrm{S} + \Phi\,\mathrm{Q} \,, \label{Smarr-extremal}
\end{equation}
and the first law of black hole mechanics
\begin{equation}
\delta \mathrm{M} = \mathrm{T}\,\delta\mathrm{S} + \Phi\,\delta \mathrm{Q} \,, \label{First-Law-extremal}
\end{equation}
are satisfied if and only if $\Phi_{int}=1$.
Finally, restoring $w$ as a free parameter, we can also verify that the Christodoulou--Ruffini mass formula~\cite{Christodoulou:1971pcn} is satisfied
\begin{equation}
\mathrm{M}^2 = \frac{\mathrm{S}}{4\pi}+\frac{\mathrm{Z}^2}{2} + \frac{\pi\,\mathrm{Z}^4}{4\,\mathrm{S}} \,, \label{Ruffini-extremal}
\end{equation}
where $\mathrm{Z} = |\pazocal{Z}| = \sqrt{\mathrm{Q}^2+\mathrm{P}^2}$\,.

\section{Cosmological generalization of the extremal solution}
\label{sec-cosmological-generalization}

As shown by Kastor and Traschen in~\cite{Kastor:1992nn}, it is possible to generalize the Majumdar--Papapetrou solution describing a configuration of multi-centered extremal Reissner--Nordstr{\"o}m black holes by introducing a \emph{positive} cosmological constant $\Lambda>0$ in the following way:\footnote{For simplicity, we write here only the electric generalization $w=1$, while the full electromagnetic case can be obtained by a duality rotation of the electromagnetic field, leaving the metric unchanged.}
\begin{equation}
\label{mp-rn-cartesian-cosmological}
ds^2 = -U^{-2}{d\tau}^2 + a^2(\tau) \,U^2 \bigl({dx}^2 + {dy}^2 + {dz}^2 \bigr) \,,\qquad 
A = U^{-1} d\tau \,,
\end{equation}
where
\begin{equation}
\label{mp-rn-cartesian-cosmological-function}
U = 1 + \sum_{i=1}^{N}\frac{m_i}{a(\tau)|\vec{x}-\vec{x}_i|}\,, \qquad a(\tau) = e^{\sqrt{\frac{\Lambda}{3}}\,\tau} \,.
\end{equation}
However, we note that under a transformation of the time coordinate 
\begin{equation}
\label{cosmological-time-transformation}
\tau \mapsto \sqrt{\frac{3}{\Lambda}}\log\sqrt{\frac{\Lambda}{3}}t \,,
\end{equation}
we obtain a solution wherein the metric and the potential can again be written in the usual Majumdar--Papapetrou form, whereas the function $U$ is now time-dependent due to an extra term which is proportional to the time coordinate $t$ and the cosmological constant $\Lambda$:
\begin{subequations}
\label{mp-br-cartesian-cosmological}
\begin{align}
ds^2 = -U^{-2}{dt}^2 \,+\, & U^2 \bigl({dx}^2 + {dy}^2 + {dz}^2 \bigr) \,, \qquad A = U^{-1} dt \,,\\
U &  = \sqrt{\frac{\Lambda}{3}}\,t + \sum_{i=1}^{N}\frac{m_i}{|\vec{x}-\vec{x}_i|} \,.
\end{align}
\end{subequations}
This represents a direct cosmological generalization, with a positive cosmological constant $\Lambda>0$, of the single extremal Reissner--Nordstr{\"o}m black hole in a Bertotti--Robinson universe~\eqref{extremal-rn-br}, obtained by setting
$N=2$, $\vec{x}_{1}=(0,0,1/B)$, $m_1=1/B$, $\vec{x}_{2}=0$, $m_2=m$ in~\eqref{mp-br-cartesian-cosmological}.

On the other hand, although the two metrics given by equations~\eqref{mp-rn-cartesian-cosmological} and~\eqref{mp-br-cartesian-cosmological} are related by the coordinate transformation~\eqref{cosmological-time-transformation}, the first metric~\eqref{mp-rn-cartesian-cosmological} does not directly contain the extremal Reissner--Nordstr{\"o}m black hole in a Bertotti--Robinson universe~\eqref{extremal-rn-br} as a direct subcase for $\Lambda\rightarrow 0$. This is due to the fact that the transformation given by equation~\eqref{cosmological-time-transformation} diverges in the limit $\Lambda\rightarrow 0$, so the two solutions are diffeomorphic only for $\Lambda\neq 0$.

Moreover, the solution written as equation~\eqref{mp-br-cartesian-cosmological} with $N=2$ may be viewed as a cosmological spacetime interpolating between the extremal Reissner--Nordstr{\"o}m black hole in Bertotti--Robinson at $t=0$ and the solution describing a pair of extremal Reissner--Nordstr{\"o}m black holes at $t=\sqrt{\frac{3}{\Lambda}}$. From this perspective, the spacetime describes an observer who, at $t=0$, is in the near-horizon region of the extremal Reissner--Nordstr{\"o}m black hole generating the Bertotti--Robinson field,\footnote{We recall that the Bertotti--Robinson spacetime is the near-horizon limit of an extremal Reissner--Nordstr{\"o}m black hole.} and, as time evolves, the cosmological expansion driven by $\Lambda>0$ causes the two full extremal charged black holes to be perceived again.

Finally, we underline that these generalizations cannot be used as bosonic backgrounds in gauged supergravity, since in that case a \emph{negative} cosmological constant is required, whereas these solutions are defined only for a \emph{positive} (nonnegative) one, due to the appearance of the cosmological constant under a square root.

\section{Conclusions}

In this article, we studied the supersymmetry of a novel class of solutions describing a possibly charged and accelerating black hole embedded in a spacetime filled with an electromagnetic field of the Bertotti--Robinson type.

In particular, we found that the only subcases satisfying the necessary condition for supersymmetry are the Bertotti--Robinson background, the extremal Reissner--Nordstr{\"o}m black hole, both of which were already known to be supersymmetric, and their combination, representing an extremal Reissner--Nordstr{\"o}m black hole embedded in Bertotti--Robinson.
For this last subcase, we explicitly obtained the corresponding Killing spinors, from which we computed the associated currents, the thermodynamic quantities and the electromagnetic charges.
Moreover, we found that this subcase also belongs to the Majumdar--Papapetrou class, which allowed us to obtain the total mass of the spacetime, and therefore to verify that the BPS bound is saturated and that the laws of black hole mechanics are satisfied; and finally to derive the generalization with a \emph{positive} cosmological constant.

There are some natural extensions of our work that can be considered:
for instance, one can study the supersymmetry properties of the general black holes in Bertotti--Robinson presented in~\cite{Ovcharenko:2025cpm} which includes, e.g., rotation.
Another possibility is the addition of NUT, swirling and Melvin parameters to the solution of~\cite{Ovcharenko:2025cpm}, as it has been done for some subcases in~\cite{Astorino:2025lih, Astorino:2026okd, Barrientos:2026shy,Ma:2026otg}, in order to establish the supersymmetry of the most general case.

The inclusion of a negative cosmological constant is by far the most compelling extension, since it would imply the generalization to the gauged supergravity realm, which represents the low-energy limit of string theories and M-theory.
Being all the supersymmetric solutions of $\pazocal{N}=2$, $D=4$ supergravity classified~\cite{Caldarelli:2003pb}, one might look for the extension in such a class.

Finally, in analogy with the corresponding spacetimes without a Bertotti--Robinson electromagnetic field, and recalling that the extremal Reissner--Nordstr{\"o}m black hole in Bertotti--Robinson~\eqref{extremal-rn-br} belongs to the Majumdar--Papapetrou (MP) class~\cite{Majumdar:1947eu, Papaetrou:1947ib}, as we stated in section~\ref{sec-BPS}, we conjecture that the addition of a NUT parameter to this spacetime, as well as an extremal rotating generalization, namely the extremal Kerr--Newman black hole in Bertotti--Robinson, should both be contained in the rotating generalization of the MP class, known as the Israel--Wilson--Perjés (IWP) class~\cite{Perjes:1971gv,Israel:1972vx}.

\acknowledgments
We are grateful to Marco Astorino and Roberto Emparan for useful discussions on the topic of this paper.
This work was partly supported by INFN.
%%%%%%%%%%%%%%%%%%%%%%%%

\appendix

\section{Tetrad and spin connection}

The tetrad used in this paper is
\begin{subequations}
\begin{align}
e^0 & = \frac{1}{\Omega}\frac{\sqrt{\Delta_r}}{r} dt\,, \\
e^1 & = \frac{1}{\Omega}\frac{r}{\sqrt{\Delta_r}} dr\,, \\
e^2 & = -\frac{1}{\Omega}\frac{r}{\sqrt{\Delta_\theta}} d\theta\,, \\
e^3 & = \frac{r\sin\theta\sqrt{\Delta_\theta}}{\Omega} d\varphi\,,
\end{align}
\end{subequations}   
which yields the following spin connection:
\begin{subequations}
\begin{align}
\omega_{t}^{\;\;01} & = \frac{\partial_r \Delta_r}{2 r^2} - \frac{\Delta_r}{r^3}-\frac{\Delta_r}{r^2}\frac{\partial_r \Omega}{\Omega}\,,\\
\omega_{t}^{\;\;02} & = \frac{\sqrt{\Delta_r\,\Delta_\theta}}{r^2}\frac{\partial_\theta \Omega}{\Omega} \,,\\
\omega_{r}^{\;\;12} & = \sqrt{\frac{\Delta_\theta}{\Delta_r}}\frac{\partial_\theta \Omega}{\Omega}\,,\\
\omega_{\theta}^{\;\;12} & = \sqrt{\frac{\Delta_r}{\Delta_\theta}}\biggl(\frac{1}{r}-\frac{\partial_r \Omega}{\Omega}\biggr)\,,\\
\omega_{\varphi}^{\;\;13} & = -\sin\theta \sqrt{\Delta_r\, \Delta_\theta}\biggl(\frac{1}{r}-\frac{\partial_r \Omega}{\Omega}\biggr) \,,\\
\omega_{\varphi}^{\;\;23} & =\frac{\sin\theta \,\partial_\theta \Delta_\theta}{2}+\cos\theta \Delta_\theta - \frac{\Delta_\theta \sin\theta \,\partial_\theta\Omega}{\Omega} \,.
\end{align}
\end{subequations}
In particular, for the extremal subcase~\eqref{extremal-rn-br}:
\begin{subequations}
\begin{align}
e^0 & = \frac{(r-m)dt}{r\sqrt{1 - 2 B(r-m)\cos\theta + B^2(r-m)^2}}\,,\\
e^1 & = \frac{rdr}{(r-m)\sqrt{1 - 2 B(r-m)\cos\theta + B^2(r-m)^2}}\,,\\
e^2 & = -\frac{rd\theta}{\sqrt{1 - 2 B(r-m)\cos\theta + B^2(r-m)^2}}\,,\\
e^3 & = \frac{r\sin\theta d\varphi}{{\sqrt{1 - 2 B(r-m)\cos\theta + B^2(r-m)^2}}}\,,
\end{align}
\end{subequations}
\begin{subequations}
\begin{align}
\omega_{t}^{\;\;01} & = \frac{m(r-m)}{r^3} + \frac{B(r-m)^2\bigl[\cos\theta - B (r-m)\bigr]}{r^2\bigl[1 - 2 B(r-m)\cos\theta + B^2(r-m)^2\bigr]}\,,\\
\omega_{t}^{\;\;02} & = \frac{B(r-m)^2\sin\theta}{r^2\bigl[1 - 2 B(r-m)\cos\theta + B^2(r-m)^2\bigr]} \,,\\
\omega_{r}^{\;\;12} & = \frac{B \sin\theta}{\bigl[1 - 2 B(r-m)\cos\theta + B^2(r-m)^2\bigr]}\,,\\
\omega_{\theta}^{\;\;12} & = \frac{(r-m)\bigl[1-B(r-2m)\cos\theta-B^2 m(r-m)\bigr]}{r\bigl[1 - 2 B(r-m)\cos\theta + B^2(r-m)^2\bigr]}\,,\\
\omega_{\varphi}^{\;\;13} & = -\frac{(r-m)\sin\theta\bigl[1-B(r-2m)\cos\theta-B^2 m(r-m)\bigr]}{r\bigl[1 - 2 B(r-m)\cos\theta + B^2(r-m)^2\bigr]} \,,\\
\omega_{\varphi}^{\;\;23} & = \cos\theta -\frac{B(r-m) \sin^2\theta}{\bigl[1 - 2 B(r-m)\cos\theta + B^2(r-m)^2\bigr]} \,.
\end{align}
\end{subequations}

\section{Majorana representation of the gamma matrices}

The (real) Majorana representation of the gamma matrices $\gamma_{a} = \bigl(\gamma_{0}, \gamma_{1}, \gamma_{2}, \gamma_{3}\bigr)$ is obtained by applying the unitary transformation
\begin{equation}
U = \frac{1}{\sqrt{2}} \begin{pmatrix} \mathds{1}_{2} & \sigma_2 \\ \sigma_2 & -\mathds{1}_{2} \end{pmatrix} \,,
\end{equation}
to the gamma matrices $\gamma'_a$ in the standard Dirac representation
\begin{equation}
\gamma_{a} = U^{-1}\gamma'_{a} U \,,
\end{equation}
thus giving
\begin{subequations}
\begin{align}
\gamma_{0}&=\begin{pmatrix} 0 & -i\sigma_2 \\ -i\sigma_2 & 0 \end{pmatrix} \,,\\
\gamma_{1}&=\begin{pmatrix}-\sigma_3 & 0\\ 0& -\sigma_3 \end{pmatrix}\,, \\
\gamma_{2}&=\begin{pmatrix} 0 & -i\sigma_2 \\ i\sigma_2 & 0  \end{pmatrix} \,,\\
\gamma_{3}&=\begin{pmatrix} \sigma_1 & 0\\ 0 & \sigma_1  \end{pmatrix} \,, \\
\gamma_{5}&=\gamma_{0}\gamma_{1}\gamma_{2}\gamma_{3}=\begin{pmatrix}i\sigma_2 & 0\\ 0& -i\sigma_2\ \end{pmatrix} \,.
\end{align}
\end{subequations}
$\sigma_{j}$ denote the standard representation of the two--dimensional Pauli matrices, i.e.
\begin{equation}
\sigma_{j} =
\begin{pmatrix}
\delta_{j3} & \delta_{j1}-i\delta_{j2} \\
\delta_{j1}+i\delta_{j2} & -\delta_{j3}
\end{pmatrix}
\,.
\end{equation}
For completeness, we write here how, given a tetrad $e^{a}_{\mu}$, one obtains the gamma matrices with a curved index
\begin{equation}
\gamma_{\mu}=e^{a}_{\mu}\gamma_{a} \,,
\end{equation}
and the gamma matrices with two curved indices
\begin{equation}
\gamma_{\mu \nu}= e^{a}_{\mu}e^{b}_{\nu}\gamma_{a b} = \frac{1}{2}e^{a}_{\mu}e^{b}_{\nu}[\gamma_{a},\gamma_{b}]\,.
\end{equation}
\section{Bilinears constructed from the Killing spinors of the extremal subcase}

In this appendix, we construct the bilinears associated with the Killing spinors of the extremal subcase~\eqref{killing-spinor-extremal} and summarize their main properties.
In $D=4$, there are only 16 independent bilinears $\bar{\epsilon}\,\Gamma \,\epsilon$, which can be obtained by taking $\Gamma$ to be a basis element of the Clifford algebra, namely $\Gamma \in \{\mathds{1},\gamma_{5},\gamma^{\mu},\gamma^{\mu}\gamma_{5},\gamma^{\mu\nu}\}$, since all higher-rank gamma products are related to these by duality relations.

For brevity, we introduce the shorthand notation
\begin{subequations}
\begin{align}
\Omega & = 1- 2 B \cos\theta (r-m) + B^2 (r-m)^2 \,,\qquad\,\,\, \Delta_r = (r-m)^2 \,,\\
U & = \bigl(1+B^2(r-m)^2\bigr)\cos\theta - 2 B (r-m)\,, \qquad  V = \bigl(1-B^2(r-m)^2\bigr)\sin\theta \,,\\
C_{\pm} & = c_{1}^{*} c_2 \pm c_{2}^{*} c_{1}\,, \qquad \qquad \qquad \qquad \qquad \qquad \, \,K_{\pm} =|c_{1}|^2\pm|c_{2}|^2 \,.
\end{align}
\end{subequations}
Accordingly, the scalar bilinear is
\begin{subequations}
\begin{align}
\label{scalar-bilinear}
\Phi_s =\sum_{n,m = 1}^2 \bar{\epsilon}_{n}\epsilon_{m} & =\frac{\sqrt{\Delta}\bigl( K_{+}-C_{+}w\bigr)}{r\, \Omega}\\
& = \frac{\Bigl[ \bigl(|c_{1}|^2+|c_{2}|^2\bigr)-\bigl(c_{1}^{*} c_2 + c_{2}^{*} c_{1}\bigr)w\Bigr]\bigl(1-\frac{m}{r}\bigr)}{\sqrt{1-2 B \cos\theta(r-m)+B^2(r-m)^2}} \,,
\end{align}
\end{subequations}
% \Rightarrow \mathrm{J}_s^2 &= - \mathrm{J}^{\mu}\mathrm{J}_{\mu} \,,\\
while the pseudoscalar bilinear vanishes,
\begin{equation}
\label{pseudo-scalar-bilinear}
\Phi_{5} =\sum_{n,m = 1}^2 i\,\bar{\epsilon}_{n}\gamma_5\epsilon_{m}=0 \,.
\end{equation}
The vector bilinear is the vector current that we computed in section~\ref{killing-spinor}, which we report again here for completeness
\begin{subequations}
\begin{align}
\label{vector-bilinear}
\mathrm{J}^{\mu}\partial_{\mu}=\sum_{n,m = 1}^2\bar{\epsilon}_{n}\gamma^{\mu}\epsilon_{m}\partial_{\mu} & = -\bigl( K_{+}-C_{+}w\bigr)\partial_{t}\\ 
& = -\Bigl[ \bigl(|c_{1}|^2+|c_{2}|^2\bigr)-\bigl(c_{1}^{*} c_2 + c_{2}^{*} c_{1}\bigr)w\Bigr]\partial_{t} \,,
\end{align}
\end{subequations}
while the axial current corresponds to the pseudovector bilinear
\begin{subequations}
\begin{align}
\label{pseudo-vector-bilinear}
\mathrm{J}^{\mu}_5\partial_{\mu} & =\sum_{n,m = 1}^2 i \,\bar{\epsilon}_{n}\gamma^{\mu}\gamma_{5}\epsilon_{m}\partial_{\mu} \\
& = \frac{\sqrt{\Delta_r}}{r^2}\Bigl[\Xi_{r}\, \partial_{r} +\,\Xi_{\theta}\, \partial_{\theta} + \,\Xi_{\varphi}\, \partial_{\varphi}\Bigr]\,,
\end{align}
\end{subequations}
where we also defined
\begin{subequations}
\begin{align}
\Xi_{r} & = \frac{\sqrt{\Delta_{r}}}{\Omega^2}\Bigl[V (C_{+}- K_{+}w) \sin\varphi -\bigl(UK_{-} - i\, VC_{-}\cos\varphi\bigr) \sqrt{1-w^2}\Bigr] \,,\\
\Xi_{\theta} & =\frac{1}{\Omega^2} \Bigl[U (C_{+}- K_{+}w) \sin\varphi + \bigl(VK_{-} + i\, U C_{-}\cos\varphi\bigr)\sqrt{1-w^2} \Bigr]\,,\\
\Xi_{\varphi} & = \frac{1}{\sin\theta}\Bigl[(C_{+}- K_{+}w)\cos\varphi- i\,C_{-} \sin\varphi \sqrt{1-w^2}\Bigr] \,.
\end{align}
\end{subequations}
Using these results we have that the following relations are satisfied:
\begin{subequations}
\begin{align}
g_{\mu \nu}\,\mathrm{J}^{\mu}\mathrm{J}_{5}^{\nu} & = 0 \,,\\
g_{\mu \nu}\,\mathrm{J}^{\mu}\mathrm{J}^{\nu} & = -g_{\mu \nu}\,\mathrm{J}_{5}^{\mu}\mathrm{J}_{5}^{\nu} =  - \Phi_s^2\,.
% g_{\mu \nu}\,\mathrm{J}^{\mu}\mathrm{J}^{\nu} & =  a_{1} \mathrm{J}_{s}^2 + a_{2} \mathrm{J}_{5}^2 + a_{3} g_{\mu \nu} \,\mathrm{J}^{\mu}_{5}\mathrm{J}_{5}^{\nu}\,,\\
% g_{\mu \nu}\,\mathrm{J}^{\mu}\mathrm{J}^{\nu} & = b_{1} \mathrm{J}_{s}^2 + b_{2} \mathrm{J}_{5}^2 \,.
\end{align}
\end{subequations}
Finally, the tensor bilinear defines an antisymmetric $2-$form,
\begin{subequations}
\begin{align}
\label{tensor-bilinear}
\frac{1}{2}f_{\mu \nu}\,dx^{\mu} \wedge dx^{\nu} & =\frac{1}{2}\sum_{n,m = 1}^2\bar{\epsilon}_{n}\gamma_{\mu \nu}\,\epsilon_{m}\,dx^{\mu} \wedge dx^{\nu} \\
\begin{split}
& = -\frac{\sqrt{1-w^2} }{r\, \Omega^3}\Bigl[ \,\Xi_{r}\,dx^{t} \wedge dx^{r} + \Delta_{r}\,\Xi_{\theta}\, dx^{t} \wedge dx^{\theta} + \Delta_{r}\sin^2\theta\,\Xi_{\varphi}\, dx^{t} \wedge dx^{\varphi} \Bigr] \\
& \quad + \frac{r\, w\sin\theta}{\Omega^3}\Bigl[\,\Xi_{\varphi}\, dx^{r} \wedge dx^{\theta} + \,\Xi_{\theta}\, dx^{\varphi} \wedge dx^{r} + \,\Xi_{r}\, dx^{\theta} \wedge dx^{\varphi} \Bigr] \,.
\end{split}
\end{align}
\end{subequations}
Moreover, a key property of the bilinears constructed from the Killing spinors is that the vector bilinear~\eqref{vector-bilinear} is, by construction, a Killing vector, timelike in this case, while the tensor bilinear~\eqref{tensor-bilinear} is an antisymmetric Killing--Yano $2-$form.
This allows us to construct a rank$-2$ symmetric Killing tensor
\begin{subequations}
\begin{align}
\label{killing-tensor}
K^{\mu \nu} \partial_{\mu}\,\partial_{\nu} & = f^{\mu}_{\;\;\rho}f^{\rho \nu}\partial_{\mu}\,\partial_{\nu} \\
\begin{split}
& = - \frac{(1-w^2)\,\Xi^2}{\Delta_{r}}\partial_{t}\,\partial_{t}  +\frac{\Delta_{r}\bigl(\,\Xi^2_{r}\,-w^2\,\Xi^2\bigr)}{r^4}\partial_{r}\,\partial_{r}\\ 
& \quad  + \frac{\bigl(\Delta_{r}\,\Xi^2_{\theta}\,-w^2\,\Xi^2\bigr)}{r^4}\partial_{\theta}\,\partial_{\theta} + \frac{\bigl(\Delta_{r}\sin^2\theta\,\Xi^2_{\varphi}\,-w^2\,\Xi^2\bigr)}{r^4 \sin^2\theta}\partial_{\varphi}\,\partial_{\varphi} + \\
& \quad + \frac{2\Delta_{r}}{r^4}\Bigl[\,\Xi_{r}\,\Xi_{\theta}\,\partial_{r}\,\partial_{\theta} + \,\Xi_{r}\,\Xi_{\varphi}\,\partial_{r}\,\partial_{\varphi} + \,\Xi_{\theta}\,\Xi_{\varphi}\,\partial_{\theta}\,\partial_{\varphi}\Bigr] \,,
\end{split}
\end{align}
\end{subequations}
where
\begin{align}
\Xi^2 = \Xi^2_{r}\,+\Delta_{r}\,\Xi^2_{\theta}\,+\Delta_{r}\sin^2\theta \,\Xi^{2}_{\varphi}\,.
\end{align}

\section{Thermodynamics and charges of the non-extremal case}
\label{app-thermo-charges-general}

In the non-extremal case, i.e.~metric~\eqref{accelerating-rn-br-metric}, the black hole possesses both an inner and an outer horizon, $r_{H}=(r_{-},r_{+})$, whose relative positions, as discussed in section~\ref{subsec-horizons}, depend on the specific values of the parameters. Since the thermodynamic quantities may vary depending on the horizon at which they are evaluated, we compute the entropy and temperature on both horizons, $r_{-} = r_0$ and $r_{+} = r_0 + \frac{2m}{1-B^2 m^2}$.
Specifically, we have that the the surface gravity $\kappa$ and the temperature $\mathrm{T}$ at the horizon, which correspond to the following quantities
\begin{equation}
\kappa = \sqrt{-\frac{1}{2} \nabla_{\mu}\xi_{\nu}\nabla^{\mu}\xi^{\nu}}\Bigg\rvert_{r=r_H} = \frac{|\Delta'(r_H)|}{2\,r_H^2} \,, \qquad \mathrm{T} = \frac{\kappa}{2\pi} \,,
\end{equation}
with $\xi =\partial_{t}$ the timelike Killing vector, results in
\begin{subequations}
\begin{align}
%\mathrm{\kappa}_{+} & = \frac{m\bigl[(1+B^2m^2)^2-4m^2\alpha^2\bigr]}{(1-B^2 m^2)^2 \, r_+^2}\,,\quad \mathrm{T}_{+} = \frac{m\bigl[(1+B^2m^2)^2-4m^2\alpha^2\bigr]}{2 \pi(1-B^2 m^2)^2 \, r_+^2}\,, \\
\mathrm{\kappa}_{+} & = \frac{m}{r_+^2}\bigl[1-(r_+ - r_0)^2\bigl(\alpha^2-B^2)\bigr]\,, \qquad\quad\,\,\, \mathrm{\kappa}_{-} = \frac{m}{\, r_0^2}\,,\\
\mathrm{T}_{+} & = \frac{m}{2 \pi\,r_+^2}\bigl[1-(r_+ - r_0)^2\bigl(\alpha^2-B^2)\bigr]\,, \qquad\mathrm{T}_{-} = \frac{m}{2\pi\, r_0^2}\,.
\end{align}
\end{subequations}
In a similar way, the horizon area $\mathrm{A}$ and the entropy $\mathrm{S}$, 
\begin{equation}
\mathrm{A} = \int_{0}^{2 \pi}d\varphi \int_{0}^{\pi} \sqrt{g_{\varphi \varphi}g_{\theta\theta}}\Bigg\rvert_{r=r_H}\! \,, \qquad \mathrm{S}=\frac{\mathrm{A}}{4} \,,
\end{equation}
are found to be:
\begin{subequations}
\begin{align}
%\mathrm{A}_{+} & = \frac{4\pi(1-B^2 m^2)^2\,r_+^2}{(1+B^2m^2)^2-4m^2\alpha^2} \,, \qquad \mathrm{A}_{-} = 4\pi \,r_0^2 \,, \\
%\mathrm{S}_{+} & = \frac{\pi(1-B^2 m^2)^2 \, r_+^2}{(1+B^2m^2)^2-4m^2\alpha^2}\,, \qquad \,\mathrm{S}_{-} = \pi\, r_0^2\,.
\mathrm{A}_{+} & = \frac{4\pi\,r_+^2}{\bigl[1-(r_+ - r_0)^2\bigl(\alpha^2-B^2)\bigr]} \,, \qquad \mathrm{A}_{-} = 4\pi \,r_0^2 \,, \\
\mathrm{S}_{+} & = \frac{\pi \, r_+^2}{\bigl[1-(r_+ - r_0)^2\bigl(\alpha^2-B^2)\bigr]}\,, \qquad \,\mathrm{S}_{-} = \pi\, r_0^2\,.
\end{align}
\end{subequations}
From these results, it follows straightforwardly that
\begin{equation}
2 \,\mathrm{T}_{-}\,\mathrm{S}_{-} = 2 \,\mathrm{T}_{+}\,\mathrm{S}_{+} = m\,.
\end{equation}
In contrast to temperature and entropy, we find that the charges are independent of the radius at which the integral is evaluated, so the following results hold in general
\begin{equation}
\label{charges-general}
\mathrm{Q} = -\frac{1}{8\pi}\int_\Sigma F^{\mu\nu} dS_{\mu \nu} = \frac{\alpha\,r_0}{B} w \,, \,\quad\,
\mathrm{P} = -\frac{1}{8\pi}\int_\Sigma {}^\star F^{\mu\nu} dS_{\mu \nu} = \frac{\alpha\,r_0}{B} \sqrt{1-w^2} \,,
\end{equation}
where we recall that ${}^\star F^{\mu\nu} = \frac{1}{2\sqrt{-g}}\varepsilon^{\mu \nu \rho \sigma} F_{\rho \sigma}$ is the Hodge dual of the Faraday tensor $F^{\mu \nu}$.

In the extremal limit~\eqref{extremal-rn-br}, for which $r_{+}=r_{-}$, these results converge to those obtained in section~\ref{sec-BPS} and section~\ref{sec-thermo-extremal}, where in particular we obtain that the temperature $\mathrm{T}$ reduces to zero, as expected.


\begin{thebibliography}{99}

%\cite{Strominger:1996sh}
\bibitem{Strominger:1996sh}
A.~Strominger and C.~Vafa,
\emph{Microscopic origin of the Bekenstein-Hawking entropy},
\href{https://doi.org/10.1016/0370-2693(96)00345-0}{\emph{Phys. Lett. B} \textbf{379} (1996), 99-104}
[\href{https://arxiv.org/abs/hep-th/9601029}{\tt arXiv:hep-th/9601029}]
[\href{https://inspirehep.net/literature/415163}{INSPIRE}]
%3779 citations counted in INSPIRE as of 21 Apr 2026

%\cite{Bekenstein:1973ur}
\bibitem{Bekenstein:1973ur}
J.~D.~Bekenstein,
\emph{Black holes and entropy},
\href{https://doi.org/10.1103/PhysRevD.7.2333}{\emph{Phys. Rev. D} \textbf{7} (1973), 2333-2346}
[\href{https://inspirehep.net/literature/80985}{INSPIRE}]
%7938 citations counted in INSPIRE as of 29 May 2026

%\cite{Hawking:1975vcx}
\bibitem{Hawking:1975vcx}
S.~W.~Hawking,
\emph{Particle Creation by Black Holes},
\href{https://doi.org/10.1007/BF02345020}{\emph{Commun. Math. Phys.} \textbf{43} (1975), 199-220}
[Erratum: \href{https://doi.org/10.1007/BF01608497}{\emph{Commun. Math. Phys.} \textbf{46} (1976), 206}]
[\href{https://inspirehep.net/literature/101338}{INSPIRE}]
%13503 citations counted in INSPIRE as of 29 May 2026

%\cite{Hawking:1994ii}
\bibitem{Hawking:1994ii}
S.~W.~Hawking, G.~T.~Horowitz and S.~F.~Ross,
\emph{Entropy, Area, and black hole pairs},
\href{https://doi.org/10.1103/PhysRevD.51.4302}{\emph{Phys. Rev. D} \textbf{51} (1995), 4302-4314}
[\href{https://arxiv.org/abs/gr-qc/9409013}{\tt arXiv:gr-qc/9409013}]
[\href{https://inspirehep.net/literature/376601}{INSPIRE}]
%446 citations counted in INSPIRE as of 21 Apr 2026

%\cite{Ferrara:1995ih}
\bibitem{Ferrara:1995ih}
S.~Ferrara, R.~Kallosh and A.~Strominger,
\emph{N = 2 extremal black holes},
\href{https://doi.org/10.1103/PhysRevD.52.R5412}{\emph{Phys. Rev. D} \textbf{52} (1995), R5412-R5416}
[\href{https://arxiv.org/abs/hep-th/9508072}{\tt arXiv:hep-th/9508072}]
[\href{https://inspirehep.net/literature/398357}{INSPIRE}]
%1016 citations counted in INSPIRE as of 21 Apr 2026

%\cite{Gauntlett:2002sc}
\bibitem{Gauntlett:2002sc}
J.~P.~Gauntlett, D.~Martelli, S.~Pakis and D.~Waldram,
\emph{G structures and wrapped NS5-branes},
\href{https://doi.org/10.1007/s00220-004-1066-y}{\emph{Commun. Math. Phys.} \textbf{247} (2004), 421-445}
[\href{https://arxiv.org/abs/hep-th/0205050}{\tt arXiv:hep-th/0205050}]
[\href{https://inspirehep.net/literature/586452}{INSPIRE}]
%348 citations counted in INSPIRE as of 21 Apr 2026

%\cite{Gauntlett:2002nw}
\bibitem{Gauntlett:2002nw}
J.~P.~Gauntlett, J.~B.~Gutowski, C.~M.~Hull, S.~Pakis and H.~S.~Reall,
\emph{All supersymmetric solutions of minimal supergravity in five-dimensions},
\href{https://doi.org/10.1088/0264-9381/20/21/005}{\emph{Class. Quant. Grav.} \textbf{20} (2003), 4587-4634}
[\href{https://arxiv.org/abs/hep-th/0209114}{\tt arXiv:hep-th/0209114}]
[\href{https://inspirehep.net/literature/594950}{INSPIRE}]
%644 citations counted in INSPIRE as of 21 Apr 2026

%\cite{Cacciatori:2007vn}
\bibitem{Cacciatori:2007vn}
S.~L.~Cacciatori, M.~M.~Caldarelli, D.~Klemm, D.~S.~Mansi and D.~Roest,
\emph{Geometry of four-dimensional Killing spinors},
\href{https://doi.org/10.1088/1126-6708/2007/07/046}{\emph{JHEP} \textbf{07} (2007), 046}
[\href{https://arxiv.org/abs/0704.0247}{\tt arXiv:0704.0247}]
[\href{https://inspirehep.net/literature/747896}{INSPIRE}]
%38 citations counted in INSPIRE as of 21 Apr 2026

%\cite{Tod:1983pm}
\bibitem{Tod:1983pm}
K.~P.~Tod,
\emph{All Metrics Admitting Supercovariantly Constant Spinors},
\href{https://doi.org/10.1016/0370-2693(83)90797-9}{\emph{Phys. Lett. B} \textbf{121} (1983), 241-244}
[\href{https://inspirehep.net/literature/191026}{INSPIRE}]
%328 citations counted in INSPIRE as of 21 Apr 2026

%\cite{Newman:1961qr}
\bibitem{Newman:1961qr}
E.~Newman and R.~Penrose,
\emph{An Approach to gravitational radiation by a method of spin coefficients},
\href{https://doi.org/10.1063/1.1724257}{\emph{J. Math. Phys.} \textbf{3} (1962), 566-578}
[\href{https://inspirehep.net/literature/8892}{INSPIRE}]
%2056 citations counted in INSPIRE as of 21 Apr 2026

%\cite{Perjes:1971gv}
\bibitem{Perjes:1971gv}
Z.~Perjés,
\emph{Solutions of the coupled Einstein Maxwell equations representing the fields of spinning sources},
\href{https://doi.org/10.1103/PhysRevLett.27.1668}{\emph{Phys. Rev. Lett.} \textbf{27} (1971), 1668}
[\href{https://inspirehep.net/literature/69361}{INSPIRE}]
%151 citations counted in INSPIRE as of 27 Apr 2026

%\cite{Israel:1972vx}
\bibitem{Israel:1972vx}
W.~Israel and G.~A.~Wilson,
\emph{A class of stationary electromagnetic vacuum fields},
\href{https://doi.org/10.1063/1.1666066}{\emph{J. Math. Phys.} \textbf{13} (1972), 865-871}
[\href{https://inspirehep.net/literature/78130}{INSPIRE}]
%207 citations counted in INSPIRE as of 27 Apr 2026

%\cite{Romans:1991nq}
\bibitem{Romans:1991nq}
L.~J.~Romans,
\emph{Supersymmetric, cold and lukewarm black holes in cosmological Einstein-Maxwell theory},
\href{https://doi.org/10.1016/0550-3213(92)90684-4}{\emph{Nucl. Phys. B} \textbf{383} (1992), 395-415}
[\href{https://arxiv.org/abs/hep-th/9203018}{\tt arXiv:hep-th/9203018}]
[\href{https://inspirehep.net/literature/323543}{INSPIRE}]
%423 citations counted in INSPIRE as of 21 Apr 2026

%\cite{Caldarelli:1998hg}
\bibitem{Caldarelli:1998hg}
M.~M.~Caldarelli and D.~Klemm,
\emph{Supersymmetry of Anti-de Sitter black holes},
\href{https://doi.org/10.1016/S0550-3213(98)00846-3}{\emph{Nucl. Phys. B} \textbf{545} (1999), 434-460}
[\href{https://arxiv.org/abs/hep-th/9808097}{\tt arXiv:hep-th/9808097}]
[\href{https://inspirehep.net/literature/474973}{INSPIRE}]
%216 citations counted in INSPIRE as of 21 Apr 2026

%\cite{Alonso-Alberca:2000zeh}
\bibitem{Alonso-Alberca:2000zeh}
N.~Alonso-Alberca, P.~Meessen and T.~Ortin,
\emph{Supersymmetry of topological Kerr-Newman-Taub-NUT-AdS space-times},
\href{https://doi.org/10.1088/0264-9381/17/14/312}{\emph{Class. Quant. Grav.} \textbf{17} (2000), 2783-2798}
[\href{https://arxiv.org/abs/hep-th/0003071}{\tt arXiv:hep-th/0003071}]
[\href{https://inspirehep.net/literature/524787}{INSPIRE}]
%103 citations counted in INSPIRE as of 21 Apr 2026

%\cite{Eatough:2013nva}
\bibitem{Eatough:2013nva}
R.~P.~Eatough, H.~Falcke, R.~Karuppusamy, K.~J.~Lee, D.~J.~Champion, E.~F.~Keane, G.~Desvignes, D.~H.~F.~M.~Schnitzeler, L.~G.~Spitler and M.~Kramer, \textit{et al.}
\emph{A strong magnetic field around the supermassive black hole at the centre of the Galaxy},
\href{https://doi.org/10.1038/nature12499}{\emph{Nature} \textbf{501} (2013), 391-394}
[\href{https://arxiv.org/abs/1308.3147}{\tt arXiv:1308.3147}]
[\href{https://inspirehep.net/literature/1247779}{INSPIRE}]
%409 citations counted in INSPIRE as of 24 Apr 2026

%\cite{You:2023dax}
\bibitem{You:2023dax}
B.~You, X.~Cao, Z.~Yan, J.~M.~Hameury, B.~Czerny, Y.~Wu, T.~Xia, M.~Sikora, S.~N.~Zhang and P.~Du, \textit{et al.}
\emph{Observations of a black hole x-ray binary indicate formation of a magnetically arrested disk},
\href{https://doi.org/10.1126/science.abo4504}{\emph{Science} \textbf{381} (2023) no.6661, abo4504}
[\href{https://arxiv.org/abs/2309.00200}{\tt arXiv:2309.00200}]
[\href{https://inspirehep.net/literature/2692933}{INSPIRE}]
%34 citations counted in INSPIRE as of 24 Apr 2026

%\cite{Bonnor:1954tis}
\bibitem{Bonnor:1954tis}
W.~B.~Bonnor,
\emph{Static Magnetic Fields in General Relativity},
\href{https://doi.org/10.1088/0370-1298/67/3/305}{\emph{Proc. Roy. Soc. Lond. A} \textbf{67} (1954) no.3, 225}
[\href{https://inspirehep.net/literature/2931009}{INSPIRE}]
%75 citations counted in INSPIRE as of 30 Apr 2026

%\cite{Melvin:1963qx}
\bibitem{Melvin:1963qx}
M.~A.~Melvin,
\emph{Pure magnetic and electric geons},
\href{https://doi.org/10.1016/0031-9163(64)90801-7}{\emph{Phys. Lett.} \textbf{8} (1964), 65-70}
[\href{https://inspirehep.net/literature/42731}{INSPIRE}]
%447 citations counted in INSPIRE as of 30 Apr 2026

%\cite{Bertotti:1959pf}
\bibitem{Bertotti:1959pf}
B.~Bertotti,
\emph{Uniform electromagnetic field in the theory of general relativity},
\href{https://doi.org/10.1103/PhysRev.116.1331}{\emph{Phys. Rev.} \textbf{116} (1959), 1331}
[\href{https://inspirehep.net/literature/42968}{INSPIRE}]
%395 citations counted in INSPIRE as of 24 Apr 2026

%\cite{Robinson:1959ev}
\bibitem{Robinson:1959ev}
I.~Robinson,
\emph{A Solution of the Maxwell-Einstein Equations},
{\emph{Bull. Acad. Pol. Sci. Ser. Sci. Math. Astron. Phys.}  \textbf{7} (1959), 351-352}
[\href{https://inspirehep.net/literature/44612}{INSPIRE}]
%198 citations counted in INSPIRE as of 24 Apr 2026

%\cite{Ernst:1976mzr}
\bibitem{Ernst:1976mzr}
F.~J.~Ernst,
\emph{Black holes in a magnetic universe},
\href{https://doi.org/10.1063/1.522781}{\emph{J. Math. Phys.} \textbf{17} (1976) no.1, 54-56}
[\href{https://inspirehep.net/literature/1673653}{INSPIRE}]
%304 citations counted in INSPIRE as of 24 Apr 2026

%\cite{Ernst:1976bsr}
\bibitem{Ernst:1976bsr}
F.~J.~Ernst and W.~J.~Wild,
\emph{Kerr black holes in a magnetic universe},
\href{https://doi.org/10.1063/1.522875}{\emph{J. Math. Phys.} \textbf{17} (1976) no.2, 182}
[\href{https://inspirehep.net/literature/2735272}{INSPIRE}]
%140 citations counted in INSPIRE as of 24 Apr 2026

%\cite{Thorne:1965xnn}
\bibitem{Thorne:1965xnn}
K.~S.~Thorne,
\emph{Absolute Stability of Melvin's Magnetic Universe},
\href{https://doi.org/10.1103/physrev.139.b244}{\emph{Phys. Rev.} \textbf{139} (1965) no.1B, B244-B254}
[\href{https://inspirehep.net/literature/3158495}{INSPIRE}]
%88 citations counted in INSPIRE as of 22 May 2026

%\cite{Gibbons:2013yq}
\bibitem{Gibbons:2013yq}
G.~W.~Gibbons, A.~H.~Mujtaba and C.~N.~Pope,
\emph{Ergoregions in Magnetised Black Hole Spacetimes},
\href{https://doi.org/10.1088/0264-9381/30/12/125008}{\emph{Class. Quant. Grav.} \textbf{30} (2013) no.12, 125008}
[\href{https://arxiv.org/abs/1301.3927}{\tt arXiv:1301.3927}]
[\href{https://inspirehep.net/literature/1211327}{INSPIRE}]
%118 citations counted in INSPIRE as of 27 Apr 2026

%\cite{Booth:2015nwa}
\bibitem{Booth:2015nwa}
I.~Booth, M.~Hunt, A.~Palomo-Lozano and H.~K.~Kunduri,
\emph{Insights from Melvin{\textendash}Kerr{\textendash}Newman spacetimes},
\href{https://doi.org/10.1088/0264-9381/32/23/235025}{\emph{Class. Quant. Grav.} \textbf{32} (2015) no.23, 235025}
[\href{https://arxiv.org/abs/1502.07388}{\tt arXiv:1502.07388}]
[\href{https://inspirehep.net/literature/1346501}{INSPIRE}]
%32 citations counted in INSPIRE as of 19 May 2026

%\cite{Podolsky:2025tle}
\bibitem{Podolsky:2025tle}
J.~Podolsky and H.~Ovcharenko,
\emph{Kerr Black Hole in a Uniform Bertotti-Robinson Magnetic Field: An Exact Solution},
\href{https://doi.org/10.1103/rfgv-ybz5}{\emph{Phys. Rev. Lett.} \textbf{135} (2025) no.18, 181401}
[\href{https://arxiv.org/abs/2507.05199}{\tt arXiv:2507.05199}]
[\href{https://inspirehep.net/literature/2942989}{INSPIRE}]
%38 citations counted in INSPIRE as of 27 Mar 2026

%\cite{Ovcharenko:2025cpm}
\bibitem{Ovcharenko:2025cpm}
H.~Ovcharenko and J.~Podolsk{\'y},
\emph{New class of rotating charged black holes with nonaligned electromagnetic field},
\href{https://doi.org/10.1103/8wkz-th6v}{\emph{Phys. Rev. D} \textbf{112} (2025) no.6, 064076}
[\href{https://arxiv.org/abs/2508.04850}{\tt arXiv:2508.04850}]
[\href{https://inspirehep.net/literature/2958289}{INSPIRE}]
%17 citations counted in INSPIRE as of 27 Mar 2026

%\cite{Ovcharenko:2025qov}
\bibitem{Ovcharenko:2025qov}
H.~Ovcharenko and J.~Podolsk{\'y},
\emph{A novel class of rotating black holes with non-aligned electromagnetic field},
\href{https://doi.org/10.1088/1742-6596/3177/1/012006}{\emph{J. Phys. Conf. Ser.} \textbf{3177} (2026) no.1, 012006}
[\href{https://arxiv.org/abs/2511.04840}{\tt arXiv:2511.04840}]
[\href{https://inspirehep.net/literature/3080905}{INSPIRE}]
%2 citations counted in INSPIRE as of 04 Jun 2026

%\cite{Ovcharenko:2026byw}
\bibitem{Ovcharenko:2026byw}
H.~Ovcharenko and J.~Podolsky, \emph{Static black holes in an external uniform electromagnetic field: Reissner-Nordstr{\"o}m accelerating in Bertotti-Robinson},
[\href{https://arxiv.org/abs/2602.15462}{\tt arXiv:2602.15462}]
[\href{https://inspirehep.net/literature/3121045}{INSPIRE}]
%3 citations counted in INSPIRE as of 27 Mar 2026

%\cite{Ovcharenko:2026pow}
\bibitem{Ovcharenko:2026pow}
H.~Ovcharenko and J.~Podolsky,
\emph{Uniqueness of stationary axisymmetric type D black holes with non-aligned electromagnetic field},
[\href{https://arxiv.org/abs/2604.13202}{\tt arXiv:2604.13202}]
[\href{https://inspirehep.net/literature/3145509}{INSPIRE}]
%1 citations counted in INSPIRE as of 05 Jun 2026

%\cite{Debever:1983pi}
\bibitem{Debever:1983pi}
R.~Debever, N.~Kamran and R.~G.~McLenaghan,
\emph{A single expression for the general solution of Einstein's vacuum and electrovac field equations with cosmological constant for Petrov type D admitting a non-singular aligned Maxwell field},
\href{https://doi.org/10.1016/0375-9601(83)90469-3}{\emph{Phys. Lett. A} \textbf{93} (1983), 399-402}
[\href{https://inspirehep.net/literature/191023}{INSPIRE}]
%10 citations counted in INSPIRE as of 24 Apr 2026

%\cite{Debever:1984yxe}
\bibitem{Debever:1984yxe}
R.~Debever, N.~Kamran and R.~G.~McLenaghan,
\emph{Exhaustive integration and a single expression for the general solution of the type D vacuum and electrovac field equations with cosmological constant for a nonsingular aligned Maxwell field},
\href{https://doi.org/10.1063/1.526386}{\emph{J. Math. Phys.} \textbf{25} (1984) no.6, 1955-1972}
[\href{https://inspirehep.net/literature/3166015}{INSPIRE}]
%19 citations counted in INSPIRE as of 08 Jun 2026


%\cite{Alekseev:1996fq}
\bibitem{Alekseev:1996fq}
G.~A.~Alekseev and A.~A.~Garcia,
\emph{Schwarzschild black hole immersed in a homogeneous electromagnetic field},
\href{https://doi.org/10.1103/PhysRevD.53.1853}{\emph{Phys. Rev. D} \textbf{53} (1996), 1853-1867}
[\href{https://inspirehep.net/literature/429965}{INSPIRE}]
%28 citations counted in INSPIRE as of 27 Mar 2026

%\cite{Alekseev:2025czq}
\bibitem{Alekseev:2025czq}
G.~A.~Alekseev,
\emph{Charged black hole accelerated by spatially homogeneous electric field of Bertotti-Robinson} ($\mathrm{AdS}^{2} \times \mathbb{S}^{2}$) \emph{space-time},
[\href{https://arxiv.org/abs/2511.06082}{\tt arXiv:2511.06082}]
[\href{https://inspirehep.net/literature/3081423}{INSPIRE}]
%3 citations counted in INSPIRE as of 27 Mar 2026

%\cite{Ferrara:1997yr}
\bibitem{Ferrara:1997yr}
S.~Ferrara,
\emph{Bertotti-Robinson geometry and supersymmetry},
{Part of Proceedings, 12th Italian Conference on General Relativity and Gravitational Physics: Rome, Italy, September 23-27, (1996), 135-147}
[\href{https://arxiv.org/abs/hep-th/9701163}{\tt arXiv:hep-th/9701163}]
[\href{https://inspirehep.net/literature/439912}{INSPIRE}]
%17 citations counted in INSPIRE as of 12 May 2026

%\cite{Andrianopoli:1996cm}
\bibitem{Andrianopoli:1996cm}
L.~Andrianopoli, M.~Bertolini, A.~Ceresole, R.~D'Auria, S.~Ferrara, P.~Fre and T.~Magri,
\emph{N = 2 supergravity and N = 2 superYang-Mills theory on general scalar manifolds: Symplectic covariance, gaugings and the momentum map},
\href{https://doi.org/10.1016/S0393-0440(97)00002-8}{\emph{J. Geom. Phys.} \textbf{23} (1997), 111-189}
[\href{https://arxiv.org/abs/hep-th/9605032}{\tt arXiv:hep-th/9605032}]
[\href{https://inspirehep.net/literature/418308}{INSPIRE}]
%585 citations counted in INSPIRE as of 27 Apr 2026

%\cite{vanNieuwenhuizen:1983wu}
\bibitem{vanNieuwenhuizen:1983wu}
P.~van Nieuwenhuizen and N.~P.~Warner,
\emph{Integrability conditions for Killing spinors},
\href{https://doi.org/10.1007/BF01223747}{\emph{Commun. Math. Phys.} \textbf{93} (1984), 277}
[\href{https://inspirehep.net/literature/193597}{INSPIRE}]
%44 citations counted in INSPIRE as of 28 Mar 2026

%\cite{Witten:1978mh}
\bibitem{Witten:1978mh}
E.~Witten and D.~I.~Olive,
\emph{Supersymmetry Algebras That Include Topological Charges},
\href{https://doi.org/10.1016/0370-2693(78)90357-X}{\emph{Phys. Lett. B} \textbf{78} (1978), 97-101}
[\href{https://inspirehep.net/literature/6563}{INSPIRE}]
%1054 citations counted in INSPIRE as of 17 Apr 2026

%\cite{Hu:2026slp}
\bibitem{Hu:2026slp}
L.~Hu, R.~G.~Cai and S.~J.~Wang,
\emph{Thermodynamics of Kerr-Bertotti-Robinson black hole},
[\href{https://arxiv.org/abs/2603.18821}{\tt arXiv:2603.18821}]
[\href{https://inspirehep.net/literature/3131582}{INSPIRE}]
%1 citations counted in INSPIRE as of 17 Apr 2026

%\cite{Majumdar:1947eu}
\bibitem{Majumdar:1947eu}
S.~D.~Majumdar,
\emph{A class of exact solutions of Einstein's field equations},
\href{https://doi.org/10.1103/PhysRev.72.390}{\emph{Phys. Rev.} \textbf{72} (1947), 390-398}
[\href{https://inspirehep.net/literature/42698}{INSPIRE}]
%530 citations counted in INSPIRE as of 29 Mar 2026

%\cite{Papaetrou:1947ib}
\bibitem{Papaetrou:1947ib}
A.~Papaetrou,
\emph{A Static solution of the equations of the gravitational field for an arbitrary charge distribution},
{\emph{Proc. Roy. Irish Acad. A} \textbf{51} (1947), 191-204}
[\href{https://inspirehep.net/literature/4126}{INSPIRE}]
%415 citations counted in INSPIRE as of 29 Mar 2026

%\cite{Maldacena:1998uz}
\bibitem{Maldacena:1998uz}
J.~M.~Maldacena, J.~Michelson and A.~Strominger,
\emph{Anti-de Sitter fragmentation},
\href{https://doi.org/10.1088/1126-6708/1999/02/011}{\emph{JHEP} \textbf{02} (1999), 011}
[\href{https://arxiv.org/abs/hep-th/9812073}{\tt arXiv:hep-th/9812073}]
[\href{https://inspirehep.net/literature/480669}{INSPIRE}]
%570 citations counted in INSPIRE as of 20 Apr 2026

%\cite{Gibbons:1982fy}
\bibitem{Gibbons:1982fy}
G.~W.~Gibbons and C.~M.~Hull,
\emph{A Bogomolny Bound for General Relativity and Solitons in N=2 Supergravity},
\href{https://doi.org/10.1016/0370-2693(82)90751-1}{\emph{Phys. Lett. B} \textbf{109} (1982), 190-194}
[\href{https://inspirehep.net/literature/11748}{INSPIRE}]
%424 citations counted in INSPIRE as of 17 Apr 2026

%\cite{Smarr:1972kt}
\bibitem{Smarr:1972kt}
L.~Smarr,
\emph{Mass formula for Kerr black holes},
\href{https://doi.org/10.1103/PhysRevLett.30.71}{\emph{Phys. Rev. Lett.} \textbf{30} (1973), 71-73}
[Erratum: \href{https://doi.org/10.1103/PhysRevLett.30.521}{\emph{Phys. Rev. Lett.} \textbf{30} (1973), 521-521}]
[\href{https://inspirehep.net/literature/73661}{INSPIRE}]
%641 citations counted in INSPIRE as of 29 May 2026

%\cite{Bardeen:1973gs}
\bibitem{Bardeen:1973gs}
J.~M.~Bardeen, B.~Carter and S.~W.~Hawking,
\emph{The Four laws of black hole mechanics},
\href{https://doi.org/10.1007/BF01645742}{\emph{Commun. Math. Phys.} \textbf{31} (1973), 161-170}
[\href{https://inspirehep.net/literature/81181}{INSPIRE}]
%3947 citations counted in INSPIRE as of 29 May 2026

%\cite{Christodoulou:1971pcn}
\bibitem{Christodoulou:1971pcn}
D.~Christodoulou and R.~Ruffini,
\emph{Reversible transformations of a charged black hole},
\href{https://doi.org/10.1103/PhysRevD.4.3552}{\emph{Phys. Rev. D} \textbf{4} (1971), 3552-3555}
[\href{https://inspirehep.net/literature/74670}{INSPIRE}]
%500 citations counted in INSPIRE as of 27 May 2026

%\cite{Kastor:1992nn}
\bibitem{Kastor:1992nn}
D.~Kastor and J.~H.~Traschen,
\emph{Cosmological multi - black hole solutions},
\href{https://doi.org/10.1103/PhysRevD.47.5370}{\emph{Phys. Rev. D} \textbf{47} (1993), 5370-5375}
[\href{https://arxiv.org/abs/hep-th/9212035v2}{\tt arXiv:hep-th/9212035}]
[\href{https://inspirehep.net/literature/341280}{INSPIRE}]
%250 citations counted in INSPIRE as of 21 May 2026

%\cite{Astorino:2025lih}
\bibitem{Astorino:2025lih}
M.~Astorino,
\emph{Black holes in the external Bertotti-Robinson-Bonnor-Melvin electromagnetic field},
\href{https://doi.org/10.1103/c5lw-53yd}{\emph{Phys. Rev. D} \textbf{112} (2025) no.10, 104077}
[\href{https://arxiv.org/abs/2508.12908}{\tt arXiv:2508.12908}]
[\href{https://inspirehep.net/literature/2961754}{INSPIRE}]
%10 citations counted in INSPIRE as of 27 Mar 2026

%\cite{Astorino:2026okd}
\bibitem{Astorino:2026okd}
M.~Astorino,
\emph{Static hairy black hole in 4D general relativity},
\href{https://doi.org/10.1103/yz86-wc3g}{\emph{Phys. Rev. D} \textbf{113} (2026) no.2, 024047}
[\href{https://arxiv.org/abs/2601.16254}{\tt arXiv:2601.16254}]
[\href{https://inspirehep.net/literature/3110808}{INSPIRE}]
%3 citations counted in INSPIRE as of 27 Mar 2026

%\cite{Barrientos:2026shy}
\bibitem{Barrientos:2026shy}
J.~Barrientos, A.~Cisterna, A.~D{\'\i}az and K.~M{\"u}ller,
\emph{From Bertotti--Robinson to Vacuum: New Exact Solutions in General Relativity via Harrison and Inversion Symmetries},
[\href{https://arxiv.org/abs/2602.17581}{\tt arXiv:2602.17581}]
[\href{https://inspirehep.net/literature/3121669}{INSPIRE}]
%2 citations counted in INSPIRE as of 27 Mar 2026

%\cite{Ma:2026otg}
\bibitem{Ma:2026otg}
L.~Ma and H.~Lu,
\emph{Demagnetizing KBR and New Ricci-flat Rotating Metric},
[\href{https://arxiv.org/abs/2605.13954}{\tt arXiv:2605.13954}]
[\href{https://inspirehep.net/literature/3155385}{INSPIRE}]
%1 citations counted in INSPIRE as of 05 Jun 2026

%\cite{Caldarelli:2003pb}
\bibitem{Caldarelli:2003pb}
M.~M.~Caldarelli and D.~Klemm,
\emph{All supersymmetric solutions of N = 2, D = 4 gauged supergravity},
\href{https://doi.org/10.1088/1126-6708/2003/09/019}{\emph{JHEP} \textbf{09} (2003), 019}
[\href{https://arxiv.org/abs/hep-th/0307022}{\tt arXiv:hep-th/0307022}]
[\href{https://inspirehep.net/literature/622533}{INSPIRE}]
%125 citations counted in INSPIRE as of 05 Jun 2026

\end{thebibliography}
\end{document}